\documentclass[journal]{IEEEtran}

\usepackage{amssymb}
\usepackage[cmex10]{amsmath}

\usepackage{tipa}
\usepackage{multirow}
\usepackage{tabularx}
\usepackage{graphicx}
\usepackage{chngpage}
\usepackage{color}
\usepackage{cite}

\def\b0{{\boldsymbol 0}}

\def\max{{\rm max}}
\def\exp{{\rm exp}}
\allowdisplaybreaks

\begin{document}

\title{Scyclone: High-Quality and Parallel-Data-Free \\ Voice Conversion Using Spectrogram and Cycle-Consistent Adversarial Networks}

\author{Masaya~Tanaka,
        Takashi~Nose,
        Aoi~Kanagaki,
        Ryohei~Shimizu,
        and Akira~Ito
}

%

\maketitle

\begin{abstract}
This paper proposes Scyclone, a high-quality voice conversion (VC) technique without parallel data training. Scyclone improves speech naturalness and speaker similarity of the converted speech by introducing CycleGAN-based spectrogram conversion with a simplified WaveRNN-based vocoder. In Scyclone, a linear spectrogram is used as the conversion features instead of vocoder parameters, which avoids quality degradation due to extraction errors in fundamental frequency and voiced/unvoiced parameters. The spectrogram of source and target speakers are modeled by modified CycleGAN networks, and the waveform is reconstructed using the simplified WaveRNN with a single Gaussian probability density function. The subjective experiments with completely unpaired training data show that Scyclone is significantly better than CycleGAN-VC2, one of the existing state-of-the-art parallel-data-free VC techniques.
\end{abstract}


\section{Introduction}
Voice conversion (VC) is a technique to convert the properties of the input speech such as speaker identity, emotion, speech naturalness, and so on, and there have been many studies in VC\cite{mohammadi2017overview}. VC has good potential for various applications such as voice impersonation in broadcasting, dubbing of movies, pronunciation training. Similar to other research fields, techniques based on deep learning have been actively proposed in recent years. One of the hot issues in VC studies is VC without parallel (paired) utterances of source and target speakers, called non-parallel or parallel-data-free VC. Since the primary and straightforward idea for the VC is the conversion of aligned data of source and target speakers (parallel VC), it is obvious that achieving good conversion quality in the parallel-data-free VC is much challenging than that in the parallel VC.

Although various approaches have been proposed to the parallel-data-free VC, the approaches based on variational autoencoders (VAEs) \cite{kigma2014auto} (e.g., \cite{hsu2016voice}), phonetic posteriorgrams (PPGs) \cite{hazen2009query} (e.g., \cite{sun2016phonetic}), cycle-consistent adversarial networks (CycleGAN) \cite{zhu2017unpaired} (e.g., \cite{fang2018high}) , or their combination (e.g., \cite{saito2018non}) seem to be more widely studied and provide better results than the other approaches. Among these approaches, the techniques based on PPGs need a large speech corpus of the target language including many speakers, which can be a limitation when applied to under-resourced languages. The VAE-based technique has also a limitation that the theory is built under the frame-wise mapping.

Recently the CycleGAN-based parallel-data-free VC technique using convolutional neural networks (CNNs) was proposed (CycleGAN-VC) \cite{kaneko2018cyclegan}. CycleGAN-VC made an impact on the parallel-data-free VC, as well as CycleGAN made on the image conversion from unpaired data \cite{zhu2017unpaired}, because the approach enables to directly optimize the sequential mapping of speech features from the source speaker to the target speaker using CNNs. Since the training of the conversion model from the unpaired speech data is more difficult than that from the paired data, the study focuses only on the conversion of spectral envelope features, which is much simpler than spectrogram conversion. They used a traditional vocoder to extract the speech parameters and reconstruct a waveform from converted parameters. Consequently, converted speech often suffers from the vocoder quality including the errors in the fundamental frequency (F0) extraction and voiced/unvoiced (V/UV) detection, which crucially degrades the speech quality. The improved version of \cite{kaneko2018cyclegan}, named CycleGAN-VC2, was also proposed \cite{kaneko2019cyclegan}, however, we found that the naturalness is still insufficient from our preliminary experiment.

In this paper, we propose a high-quality and parallel-data-free VC technique, named {\it Scyclone}, by introducing the conversion of unpaired spectrograms using CycleGAN with a neural vocoder based on a simplified version of WaveRNN \cite{kalchbrenner2018efficient}. By using the spectrogram instead of vocoder parameters as the conversion feature, we avoid the inherent problems in the conventional vocoder parameter extraction. To train the accurate conversion model from the unpaired spectrograms, we change the network architecture from the encoder-decoder model used in the original CycleGAN \cite{zhu2017unpaired} and CycleGAN-based VC techniques \cite{fang2018high,kaneko2018cyclegan,kaneko2019cyclegan} to the generator without encoder-decoder architecture. The spectral normalization with the hinge loss is also employed in discriminators to stabilize the network training. In the vocoder part, we simplify the structure of WaveRNN by introducing a Gaussian loss which is shown to be effective in text-to-speech synthesis \cite{ping2018clarinet}. To demonstrate the performance of our Scyclone, we compare Scyclone with CycleGAN-VC2, one of the state-of-the-art parallel-data-free VC techniques, through subjective experiments.

\section{CycleGAN-based spectrogram conversion with modified networks}

The basic theory of the feature conversion phase of Scyclone is the same as that of CycleGAN and CyleGAN-based VC techniques \cite{fang2018high,kaneko2018cyclegan,kaneko2019cyclegan}. We change several parts to improve training and conversion performance of Scyclone. This section describes modification in the CycleGAN-based feature conversion from the conventional CycleGAN-based VC techniques.

\subsection{Unified Conversion of Speech Features Using CycleGAN with Linear Spectogram}

The use of the traditional vocoder such as WORLD \cite{morise2016world} an advantage that the speech waveform is decomposed into the spectral envelope and excitation features, and conversion task becomes easier than that using the waveform itself or spectrogram. However, there is also a drawback that converted speech suffers from vocoder quality and V/UV errors. Another problem is that the independent conversion of respective features makes it difficult to take account of the relation among the features in conversion. In the previous studies \cite{fang2018high,kaneko2018cyclegan,kaneko2019cyclegan}, only the mel-cepstral coefficients are converted using CycleGAN, and log F0 values are linearly converted without conversion of band aperiodicity. Since the spectral envelope features include the effect of excitation, ignoring the dependency can affect the conversion performance.

The spectrogram includes both spectral and excitation features as a unified form and is widely used in recent high-fidelity text-to-speech (TTS) techniques (e.g., Tacotron2 \cite{shen2018natural}). Hence, we employ the spectrogram in the CycleGAN-based VC, which avoids the degradation caused by the parametric vocoding. In \cite{shen2018natural}, mel spectrogram is used as the target feature. However, from a preliminary experiment, we found that the low-dimensional linear spectrogram gives a better result than the mel spectrogram as the input of the following WaveRNN-based vocoder. Therefore, we use a linear spectrogram through Scyclone.

\subsection{The Use of Sectral Normalization with Hinge Function for Adversarial Loss}

Since the spectrogram conversion is a more difficult task than the spectral envelope conversion, we carefully choose a loss function for CycleGAN. In the original CycleGAN and CycleGAN-based VC techniques, the Jensen-Shannon (JS) divergence is used as the loss function with instance normalization \cite{ulyanov2016instance}. However, in the GAN training, it is known that the discriminator tends to be more quickly optimized and the training of the generator often fails. One of the reasons is that the loss of the discriminator continues to decrease theoretically during the training when the JS divergence is used.

To alleviate the problem, we employ the spectral normalization \cite{miyato2018spectral} with the hinge loss. The spectral normalization is one of the normalization techniques to stabilize the training of the discriminator. The hinge loss function is also used in the energy-based GAN \cite{zhao2016energy}. In our case, the loss functions $L_G$ and $L_D$ to be minimized for the generator $G$ and the discriminator $D$ are given by
\begin{align}
L_G =& \mathbb{E}_{x \sim p_{\rm data}(x)} [\max (0, -D_x (G_{yx}(y)))] \nonumber \\
     & + \mathbb{E}_{y \sim p_{\rm data}(y)} [\max (0, -D_y (G_{xy}(x)))] \nonumber \\
     & + \lambda_{\rm cy} \mathbb{E}_{x \sim p_{\rm data}(x)} [||G_{yx}(G_{xy}(x))-x||_1] \nonumber \\
     & + \lambda_{\rm cy} \mathbb{E}_{y \sim p_{\rm data}(y)} [||G_{xy}(G_{yx}(y))-y||_1] \nonumber \\
     & + \lambda_{\rm id} \mathbb{E}_{x \sim p_{\rm data}(x)} [||G_{yx}(x)-x||_1] \nonumber \\
     & + \lambda_{\rm id} \mathbb{E}_{y \sim p_{\rm data}(y)} [||G_{xy}(y)-y||_1] \label{eq:lg}
\end{align}
and
\begin{align}
L_D =& \mathbb{E}_{x \sim p_{\rm data}(x)} [\max (0, m-D_x (x)] \nonumber \\
     & + \mathbb{E}_{y \sim p_{\rm data}(y)} [\max (0, m-D_y (y)] \nonumber \\
     & + \mathbb{E}_{x \sim p_{\rm data}(x)} [\max (0, m+D_x (G_{yx}(y)))] \nonumber \\
     & + \mathbb{E}_{y \sim p_{\rm data}(y)} [\max (0, m+D_y (G_{xy}(x)))], \label{eq:ld}
\end{align}
respectively. $D_x(\cdot)$ and $D_y(\cdot)$ are the outputs of the discriminators for source and target speakers when given input features, respectively. $G_{xy}(\cdot)$ and $G_{yx}(\cdot)$ are the spectrogram mapping functions from source to target speakers and target to source speakers, respectively. $m$ is a parameter of the hinge loss and is set to $1.0$ in \cite{miyato2018spectral} and $0.5$ in our experiments.
By employing the hinge loss, we expect that the parameter update of discriminator stops in the early step of the training, which leads to avoid the excessive optimization of discriminator. In Eq.\,(\ref{eq:lg}), the first and the second terms are adversarial losses, the third and the forth terms are the cycle losses, and the fifth and the sixth terms are the identity mapping losses. $\lambda_{\rm cy}$ and $\lambda_{\rm id}$ are the weights for the cycle losses and the identity losses, respectively.

\subsection{Non-Encoder-Decoder Network Architecture}

Both networks of generator and discriminator mainly consist of residual blocks with one-dimensional convolutional layers. The frequency bins of the spectrogram are represented by channels in each layer, and the convolution is performed using time-domain filters. The original CycleGAN and the conventional CycleGAN-based VC techniques use the encoder-decoder model with downsampling and upsampling. This operation enables capturing a wide range of input patterns and make highly abstracted features. However, we think that such high-level abstraction increases the risk of destroying linguistic information and the time structure of input speech. A similar effect is reported in \cite{tanaka2019wavecyclegan2} as aliasing in the converted waveform. Therefore, we remove the encoder-decoder architecture and set the stride to 1 in all convolution layers of Scyclone.

\begin{figure}[t]
  \begin{center}
    \includegraphics[height=6cm]{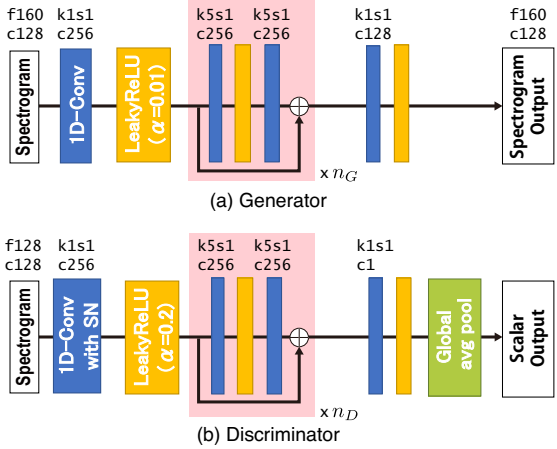}\\
    \caption{Network architectures of the generator and the discriminator of Scyclone. f, c, k, s, and SN represent the number of frames, the number of channels, the kernel size, the stride size, and the spectral normalization, respectively.}
    \label{fig:g_and_d}
  \end{center}
\end{figure}

Figure\,\ref{fig:g_and_d} shows the network architectures of the generator and the discriminator of Scyclone.
The generator doubles the number of channels\footnote{Each channel corresponds to its own bin in input and output layers.} in the first convolutional layer, and goes through the $n_G$ residual blocks consisting of two convolutional layers with a kernel size five. We used leaky rectified linear units (ReLUs) \cite{maas2013rectifier} as an activation function for both generator and discriminator. Finally, the generator halves the number of channels in the final convolutional layer. Sixteen frames of both edges of the output are discarded to ignore the effect of zero padding in the generator, and the rest of the frames are fed to the discriminator. The discriminator has a similar architecture to that of the generator. $n_D$ is the number of residual blocks in the discriminator. After the residual blocks, the number of channels is reduced to 1 in the final convolutional layer. Finally, the global average pooling \cite{lin2013network} is applied and the scalar value is outputted. We add small Gaussian noise following $\mathcal N(0, 0.01)$ to the input of the discriminator to fix the instability and vanishing gradients issues \cite{martin2017towards}. In this study, we set $n_G$ and $n_D$ to 7 and 6, respectively, based on preliminary experiments.

\section{Waveform reconstruction using simplified WaveRNN-based vocoder with Gaussian loss}

WaveRNN \cite{kalchbrenner2018efficient} is one of efficient sequential generative models for the TTS and is used as a neural vocoder as well as the WaveNet vocoder \cite{tamamori2017speaker}. The WaveRNN consists of a single-layer recurrent neural network (RNN) with a dual softmax layer that is designed to efficiently predict 16-bit raw audio samples by splitting the state of the RNN into 8 coarse bits and 8 fine bits. However, such a two-step operation increases prediction time per sample. In contrast, the joint probability of the wave samples is modeled using a single Gaussian probability density function (pdf) in \cite{ping2018clarinet}.

\begin{figure}[t]
  \begin{center}
    \includegraphics[width=1.0\hsize]{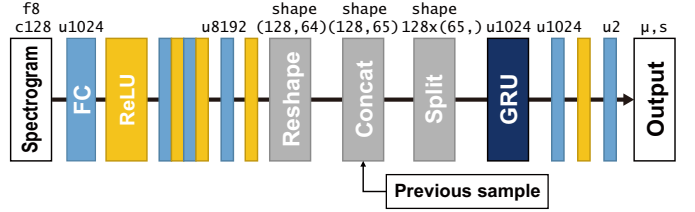}\\
    \caption{Network architecture of our WaveRNN-based vocoder. f, c, u, and FC represent the number of frames, the number of channels, the number of units, and a fully-connected layer, respectively.}
    \label{fig:wavernn}
  \end{center}
\end{figure}

Figure\,\ref{fig:wavernn} shows the network architecture of our vocoder. The network consists of four fully-connected layers with the ReLU activation functions, a single layer gated recurrent unit (GRU) \cite{cho2014learning}, and two fully-connected layers with a single ReLU activation function. As the input, eight successive frames of spectrogram with 128 frequency bins are concatenated and are used as the conditions of 128 waveform sample points that correspond to the center frame of the input spectrogram. The input spectrogram is upsampled using four fully-connected layers. The dimension of the input layer is 128 $\times$ 8 frames (1,024 units), and the dimension of the output layer is 64 $\times$ 128 sample points (8,192 units), where each sample point is conditioned on the corresponding 64-dimensional vector. The value of the previous sample point is added as the input of GRU. As the values, we use the ground-truth waveform samples in the training phase and predicted values in the inference phase.

Similarly to \cite{ping2018clarinet}, we assume that output pdf of each waveform sample is modeled by a single Gaussian pdf with a mean $\mu$ and a variance $\sigma^2$. The network outputs $\mu$ and $s = \log (\sigma)$. In the prediction, each sample point of the waveform is sampled from
\begin{equation}
p(x|\mu, s) = \frac{1}{\sqrt{2 \pi \, \exp(2s)}} \, \exp \left( - \frac{(x-\mu)^2}{\exp(2s)} \right), \label{eq:px}
\end{equation}
where $x$ represents the variable for the waveform samples. For the parameter optimization, the negative log likelihood of Eq.\,\ref{eq:px} is used as the loss function, $L_W$, given by
\begin{equation}
L_W = \frac{1}{2} \left( \log(2 \pi) + 2s + \frac{(x_t-\mu)^2}{\exp(2s)} \right), \label{eq:lw}
\end{equation}
where $x_t$ is the observed waveform sample at time $t$. We show this simplified WaveRNN-based vocoder can generate high-fidelity speech in the following experiments.

\section{Experiments}

\subsection{Experimental Conditions}

We compared Scyclone with CycleGAN-VC2 \cite{kaneko2019cyclegan} using large amount of Japanese speech data of two professional female speakers, F009 used in \cite{kawai2004ximera} and Ayanami. Ayanami is a voice impersonator of a famous Japanese anime character, Rei Ayanami\footnote{\tt https://en.wikipedia.org/wiki/Rei\_Ayanami}. 4,999 utterances of F009 and 4,973 utterances of Ayanami were used for the training. Those utterances are completely unpaired. 53 parallel utterances of the speakers were used for the evaluation. All of the utterances were used after downsampling to 16 kHz. In Scyclone, the spectrogram was calculated using a 254-point Hanning window with a 128-point shift. In the training, successive 160 frames of the spectrogram were chosen from the training data and were used as the input and the output of the generator of CycleGAN. The first and the last 16 frames were discarded and 128 frames were input to the discriminator.
$\lambda_{cy}$ and $\lambda_{id}$ were set to 10 and 1 in Eq.\,\ref{eq:lg}. Adam \cite{kingma2014adam} was used as the optimizer for both CycleGAN and WaveRNN. As the hyperparameters of Adam, $\alpha = 2.0 \times 10^{-4}$, $\beta_1 = 0.5$, and $\beta_2 = 0.999$ were used for CycleGAN, $\alpha = 1.0 \time 10^{-4}$, $\beta_1 = 0.5$, and $\beta_2 = 0.999$ were used for WaveRNN. The sizes of the mini-batches were 64 and 160 for CycleGAN and WaveRNN, respectively. For CycleGAN-VC2, we used the same conditions in feature extraction and network setting as \cite{kaneko2019cyclegan}.

\subsection{Subjective Evaluation}

We evaluated the naturalness and similarity of the converted speech samples with the mean opinion score (MOS) tests. In the speaker similarity test, the natural speech samples of the source speakers were also evaluated as the reference. This is important because the conversion task would be easy when the voice characteristics of the source and target speakers are very close to each other. Subjects were nine Japanese native speakers, and ten sentences for each subject were randomly chosen from the 53 test sentences\footnote{Some of the speech samples used in the MOS tests are available at the following URL: {\tt https://bit.ly/2NFvLhk}}. The subjects listened to the speech samples and evaluated those speech naturalness and speaker similarity to the target speaker on a five-point scale: ``1'' for bad, ``2'' for poor, ``3'' for fair, ``4'' for good, and ``5'' for excellent. 

\begin{table}[t]
  \centering
  \caption{MOS on speech naturalness}
  \label{tab:mos}
  \begin{tabular}{|c|c|c|} \hline
    Target speaker & Ayanami & F009 \\ \hline\hline
    Target (natural) & 4.59 $\pm$ 0.12  & 4.76 $\pm$ 0.10 \\ \hline
    Target (vocoded) & 4.63 $\pm$ 0.10 & 4.68 $\pm$ 0.11 \\ \hline\hline
    CycleGAN-VC2 & 1.80 $\pm$ 0.14 & 1.87 $\pm$ 0.17 \\ \hline
    Scyclone & {\bf 3.92 $\pm$ 0.17} & \bf{3.36 $\pm$ 0.20} \\ \hline
  \end{tabular}
\end{table}
\begin{table}[t]
  \centering
  \caption{MOS on similarity to the target speaker}
  \label{tab:dmos}
  \begin{tabular}{|c|c|c|} \hline
    Target speaker & Ayanami & F009 \\ \hline\hline
    Source (natural) & 1.00 $\pm$ 0.00 & 1.06 $\pm$ 0.07 \\ \hline
    CycleGAN-VC2 & 2.81 $\pm$ 0.16 & 3.20 $\pm$ 0.17 \\ \hline
    Scyclone & {\bf 4.39 $\pm$ 0.14} & {\bf 4.49 $\pm$ 0.14} \\ \hline
  \end{tabular}
\end{table}

Tables\,\ref{tab:mos} and \ref{tab:dmos} show the results of the subjective evaluation on speech naturalness and speaker similarity, respectively. From Table\,\ref{tab:mos}, we found that the proposed simplified WaveRNN-based neural vocoder using a single Gaussian loss achieves high-fidelity for both target speakers. As for the conversion performance, the naturalness of the speech samples with CycleGAN-VC2 is still low. In contrast, Scyclone gives significantly higher naturalness scores compared to CycleGAN-VC2 in both target speakers. As for the speaker similarity, the scores show that the source speakers have completely difference speaker characteristics from the target speakers. Under such a condition, Scyclone outperforms CycleGAN-VC2 in the conversion performance of the speaker identity.

\section{Conclusions}
This paper has proposed Scyclone, a parallel-data-free VC technique using CycleGAN-based spectrogram conversion and a simplified WaveRNN-based neural vocoder with a Gaussian loss. To improve the modeling and conversion performance in CycleGAN, the network was modified in which non-encoder-decoder architecture was employed with the spectral normalization. Experiments were conducted under the condition of completely unpaired training data. The subjective evaluation results have shown the superiority of Scyclone to the state-of-the-art parallel-data-free VC, CycleGAN-VC2. More detailed description and evaluation of Scylone will be presented in our next article.

\section*{Acknowledgments}
A part of this work was supported by the JST COI Grant Number JPMJCE1303 and the JSPS Grant-in-Aid for Scientific Research JP17H00823.



\bibliography{refs}
\bibliographystyle{IEEEtran}

\end{document}